\documentclass[mathleft]{an}
\usepackage{graphicx}
\usepackage{times}
\usepackage{color} 
\begin{document}

\Pagespan{789}{}
\Yearpublication{2010}%
\Yearsubmission{2010}%
\Month{11}%
\Volume{999}%
\Issue{88}%

\title{German Science Center for the Solar Dynamics Observatory}

\author{Y. Saidi\inst{1,2}
\and  R. Burston\inst{1}
\and  H. Moradi\inst{1}
\and  L. Gizon\inst{1}\fnmsep\thanks{Corresponding author:
  \email{gizon@mps.mpg.de}\newline}
}
\titlerunning{German Science Center for SDO}
\authorrunning{Y. Saidi et al.}
\institute{
Max Planck Institute for Solar System Research, Max-Planck-Str. 2,
37191 Katlenburg-Lindau, Germany
\and
Institut d'Astrophysique Spatiale (IAS), Centre universitaire d'Orsay, 91405 Orsay Cedex, France}

\received{xx xxx 2010}
\accepted{xx xxx 2010}
\publonline{later}


\abstract{
A data and computation center for helioseismology has been set up at the Max Planck Institute for Solar System Research in Germany to prepare for the SDO mission. Here we present the system infrastructure and the scientific aims of this project, which is funded through grants from the German Aerospace Center and the European Research Council. }

\maketitle

\section{Introduction}
The Solar Dynamics Observatory (SDO), launched in \linebreak February of this year, is the most important helioseismology mission of the coming decade, forming part of NASA's \linebreak Living With a Star program. The main objective of SDO is to better understand solar variability and its inevitable \linebreak impacts on the Earth and near-Earth environment. To \linebreak achieve this goal, SDO carries a payload consisting \linebreak of 3 scientific instruments: the Helioseismic and Magnetic Imager (HMI), the Atmospheric Imaging Assembly (AIA), and the Extreme ultraviolet Variability Experiment (EVE). SDO has a continuous downlink data rate of 130~Mbits per second.

Headquartered at Stanford University, the Joint Science Operations Center (JSOC) is in charge of collecting, processing, and archiving the HMI and AIA data. JSOC will make the data available to the scientific community (as well as basic science data products). The expected data volume of  $1.5$~Tbytes per day will make the analysis and processing of SDO data extremely challenging.

The German Science Center for SDO, hosted by the Max Planck Institute for Solar System Research (MPS), is a scientific IT infrastructure built around two projects:
\begin{itemize}
\item[--]
The German Data Center for SDO (GDC-SDO), funded by the German Aerospace Center (DLR), will collect, manage, and store all the calibrated HMI data as well as selected AIA data sets. It will be a master European distribution center for HMI data.
\item[--]
The Seismic Imaging of the Solar Interior (SISI) project, supported by a European Research Council (ERC) Starting Grant, will deliver specific helioseismic analyses of HMI data.
\end{itemize}

Here we present the scientific aims and the system infrastructure of the German Science Center for SDO.

\section{Scientific objectives}

\subsection{SDO/HMI observables}

HMI is the most important experiment for helioseismology science. The instrument will observe the full solar disk with a resolution of 1 arcsec. HMI comprises two CCD cameras used to acquire $4096^2$ pixels images every 45~s and 90~s. The primary HMI observables are:
\begin{itemize}
  \item[--] Dopplergrams, line-of-sight magnetograms, and continuum intensity images, with a cadence of 45~s.
  \item[--] Vector magnetograms at a lower cadence.
\end{itemize}

Compared to its predecessor, the Michelson Doppler \linebreak Imager (MDI) on SOHO, HMI will offer higher spatial resolution, a (nearly) continuous coverage of the entire solar disk, and a higher time cadence. These improvements, will allow the study of quiet and active regions all year round. The full-disk high-resolution will permit the helioseismic analysis of regions closer to the solar limb. This will benefit the study of the evolution of active regions as they rotate across the entire solar disk, and the study of the solar dynamics (e.g., the meridional flow) at higher heliographic latitudes.

JSOC will process HMI data and deliver standard helioseismology data products, as described by, e.g.,  Kosovichev (2007).
The German Science Center for SDO will focus on more specialized analyses of HMI data, as explained below.\footnote{HMI and AIA applications other than helioseismology are also planned at the German Science Center, but they are not discussed in this paper.}

\subsection{Three dimensional imaging of the solar interior}

One objective of the SISI project is to search for the root causes of solar magnetic activity by establishing physical relationships between internal solar properties and the various components of magnetic activity in the solar atmosphere. Under the assumption that subsurface \linebreak inhomogeneities are weak, it is possible to write a linear relationship between helioseismic travel times and internal solar properties. This linear relationship is given by three-dimensional sensitivity kernel functions computed using the first Born approximation, e.g., (Birch \& Gizon 2007). Based on a linear inversion procedure known as optimally localized averaging, e.g, (Jackiewicz, Gizon \& Birch 2008), we are planning to invert SDO/HMI travel times to infer vector flows and two thermodynamics quantities in the solar interior, in three dimensions. 
Developed using MATLAB, the inversion code can be parallelized into independent tasks where no intercommunication between processors is \linebreak  required. We estimate that an inversion for flows in the top 20~Mm below the photosphere, will require about $10^6$ processors ($2.2$~GHz) over 1 hour of computation.

\subsection{Seismology of active regions and sunspots}

Another objective of SISI is to infer the subsurface structure of active regions and sunspots using both SDO/HMI data and numerical simulations of the propagation of magneto-acoustic waves. Unlike in the case of convective flows, the effect of the magnetic field on solar waves cannot be considered to be small near the solar surface. In sunspots and active regions, the nature of waves is altered as they partially convert into magneto-acoustic waves. Such interactions cannot be studied using the first-order Born approximation and standard linear inversions of travel times. The only way to understand what is happening is to use numerical simulations. In particular, we are using the SLiM code (Cameron, Gizon \& Duvall 2008) to compute the propagation of small-amplitude waves through three-dimensional magnetized solar atmospheres. 

The SLiM code is an MPI based Fortran code, that is used to compute the seismic signature of model sunspots (Cameron, Gizon \& Schunker 2010). Typically, the simulation of $2.8$~hours of solar time for a $200\times200\times1098$ simulation box requires about 156 processors (2.7~GHz) over 48 hours. Many such simulations will be needed to interpret SDO/HMI observations and retrieve the subsurface structure of solar active regions.

\section{German science center for SDO}
\subsection{SDO data distribution}
 In order to optimize SDO data flow from the USA to Europe, two European master sites have been selected to automatically receive the data over the internet as is becomes available at the JSOC. One of them, GDC-SDO will automatically receive science-ready HMI data, plus low-cadence AIA data sets. In addition, LTO-4 tapes may be used to transfer shorter-term full-cadence AIA data sets. The total data volume will amount to about 70~Tbytes per year. The entire HMI data series will be archived locally at the GDC-SDO, thus it could register as a Virtual Solar Observatory (VSO) provider for those data in order to enhance the dissemination of SDO data to larger scientific community.

\subsection{Hardware}
The hardware system of the German Science Center for SDO has been dimensioned to provide adequate storage resources to save the entire science-ready HMI data and selected AIA datasets, and to offer sufficient and dedicated computing resources for data processing and numerical simulations to address the science objectives. The currently deployed hardware (figure 1) comprises several components:

\begin{figure}
\includegraphics[height = 75mm]{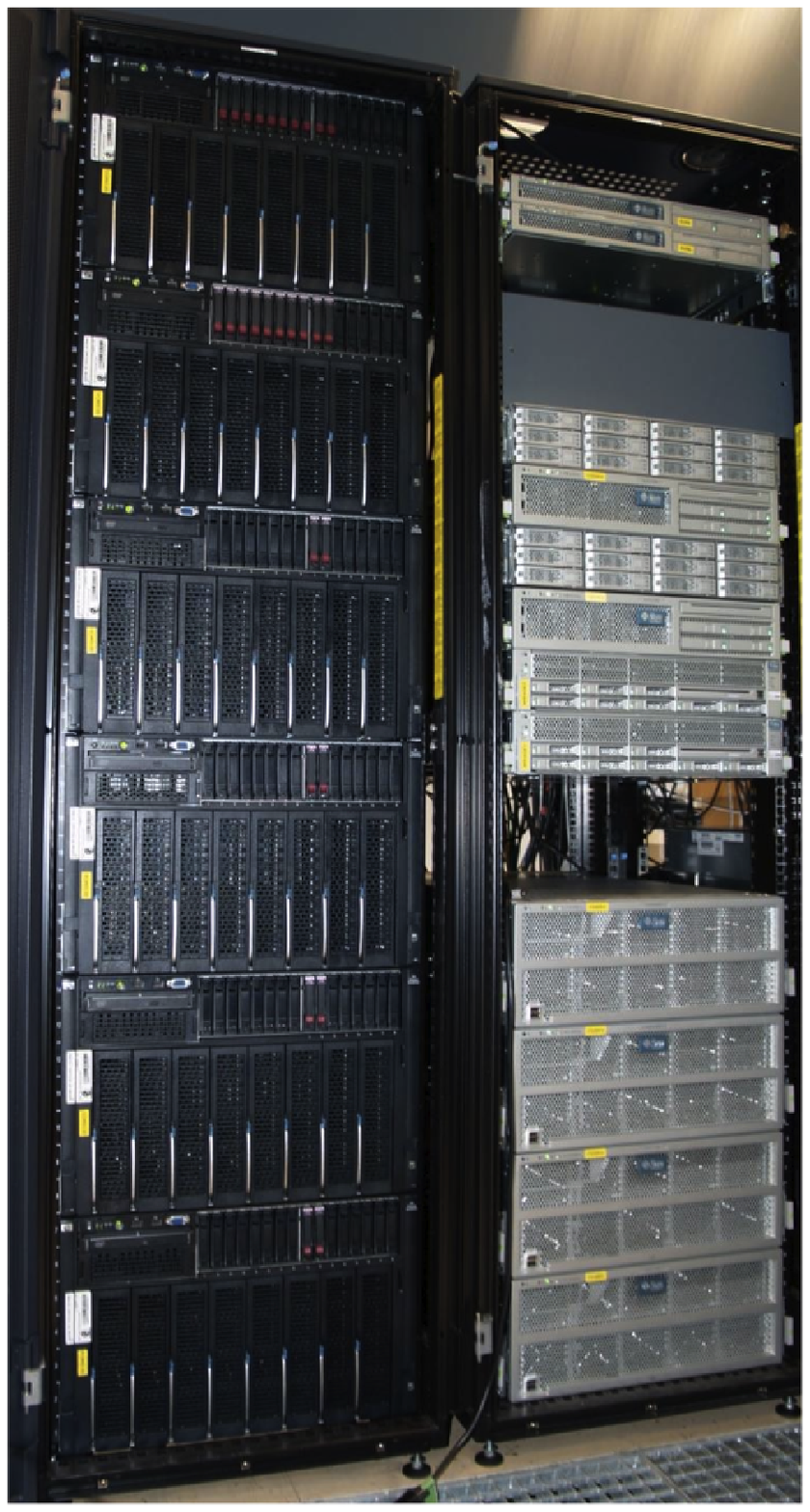}
\includegraphics[height = 75mm]{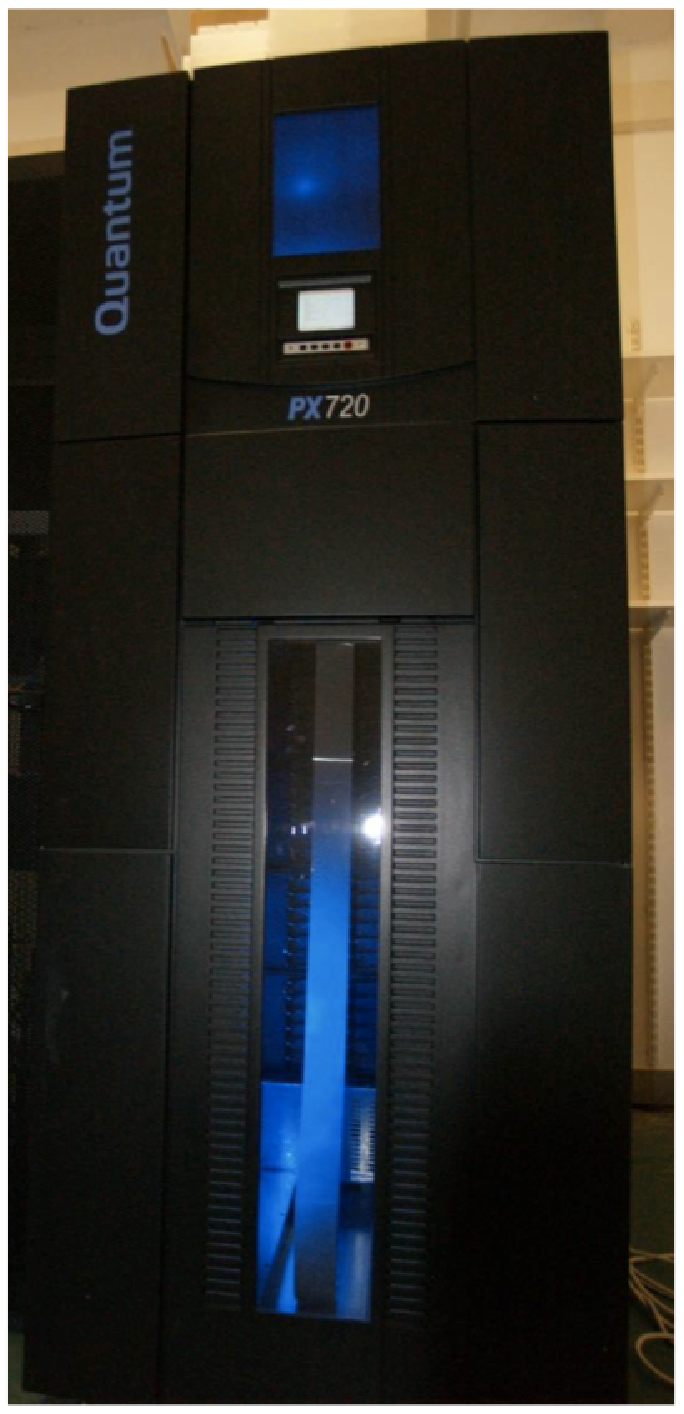}
\caption{Hardware. Left panel: the rack on the left holds the SISI compute cluster (six HP Proliant DL785GS), the rack on the right holds the GDC-SDO storage system (two pairs of Sun Fire x4500 and x4540) and interactive machines. Right panel: Tape library (Quantum PX720).}
\label{label1}
\end{figure}

\begin{figure*}
\includegraphics[width=1.0\textwidth]{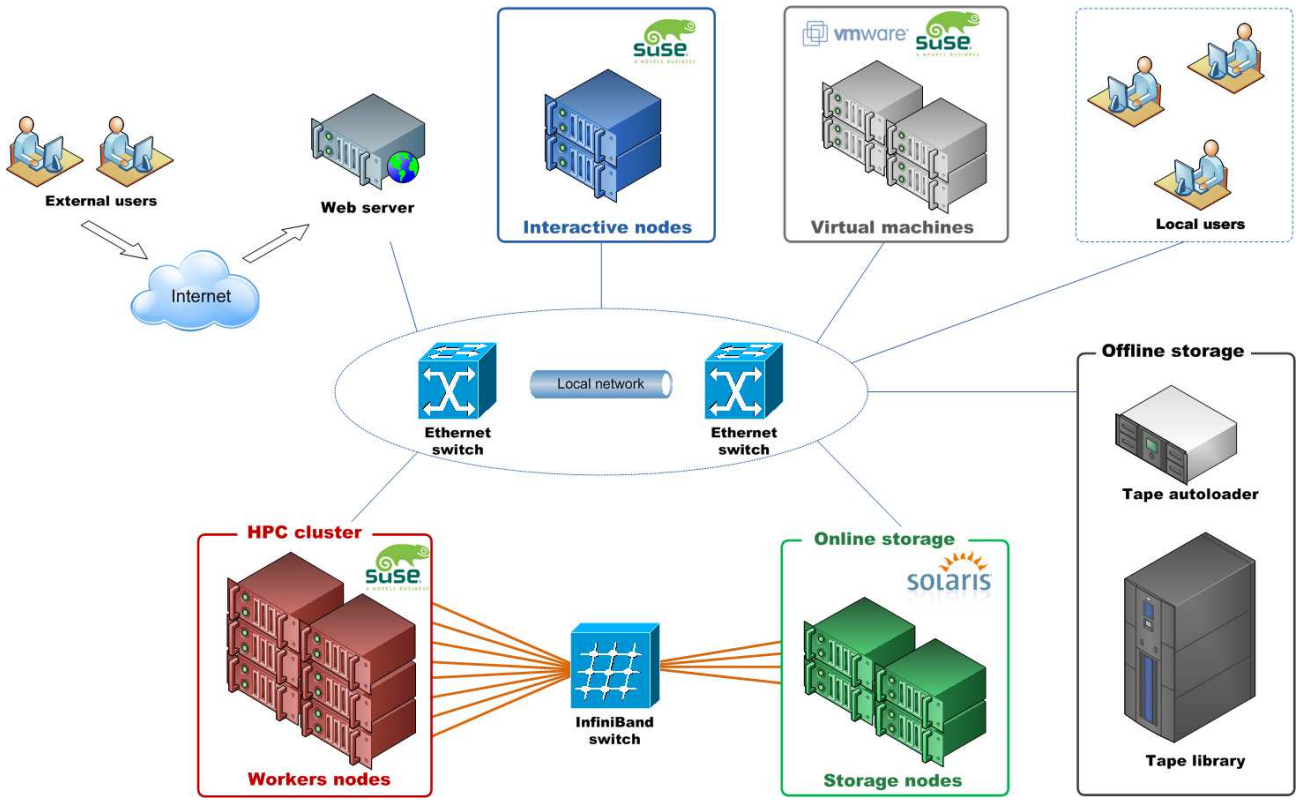}
\caption{Interconnection between various hardware components of the German Science Center for SDO.}
\label{label2}
\end{figure*}

\begin{figure*}
\includegraphics[width=1.0\textwidth]{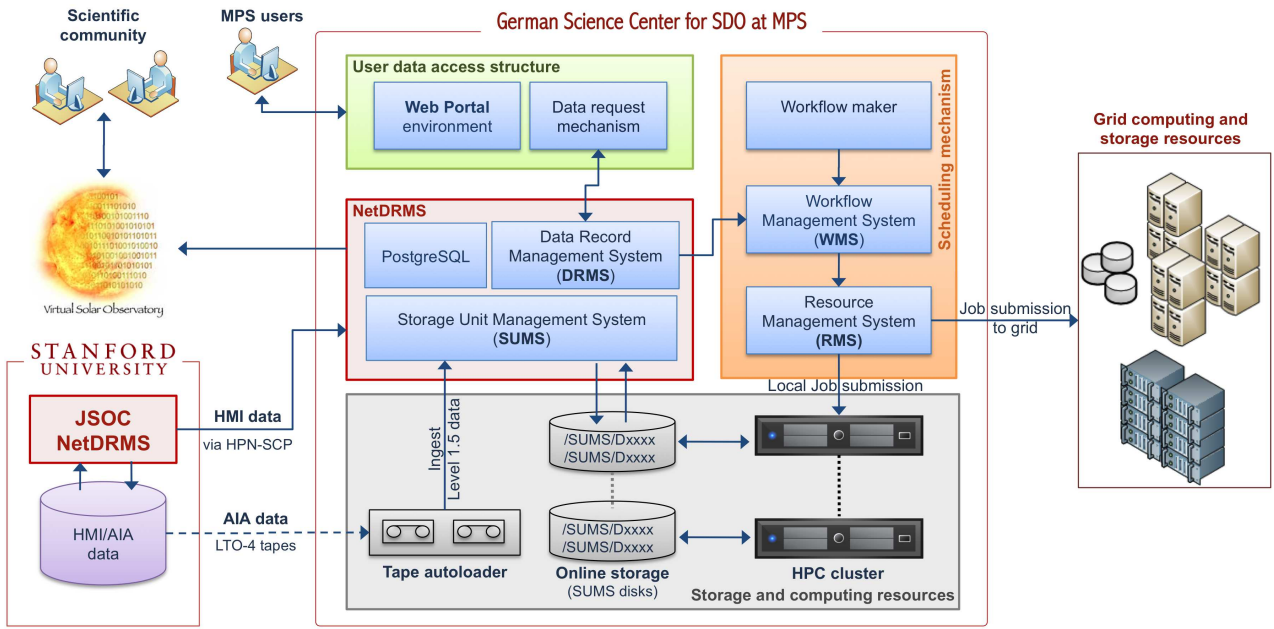}
\caption{This synthetic representation shows data movement and access, as well as software interaction with local hardware components. JSOC (USA) will be made HMI and AIA data accessible (as they became available) to the mirror sites. Using NetDRMS mechanism the German Science Center will fetch all HMI and selected AIA data from JSOC and store them locally. Using the same mechanism, we will provide all the available data collection to the scientific community through a Virtual Solar Observatory (VSO).  Locally, the data are accessible via restricted users web-portal, which helps also to create helioseismic workflows and monitor the execution of the scientific modules on local or geographically distant resources.}
\label{label3}
\end{figure*}

\begin{itemize}
\item[--]
High Performance Compute (HPC) cluster, comprising 6 identical compute nodes, with a total of 192 processing cores, 1.5~Tbytes of memory, and 2.5~Tbytes of disk space. This system has a theoretical peak performance of approximately 1.7~Tflops.
\item[--]
High performance fabric solution based on InfiniBand (IB) technology. The IB switch (figure 2) used can operate up to 20~Gbits/s data rate for nodes connectivity. In addition, it provides fully non-blocking 480~Gbits/s cross-sectional bandwidth with less than 200~ns of port-to-port latency.
\item[--]
Online storage platform, based on a series of 4 servers totalling 192~Tbytes of disk space. This storage capacity will be sufficient for roughly 2 years of selected SDO data collection.
\item[--]
Archiving solution which consists of an enterprise tape library (figure 1) with 200 DLT-S4 reserved slots providing 160~Tbytes of archiving capacity, and a tape autoloader (figure 2) with 24 slots and 2 LTO-4 drives.
\end{itemize}

Furthermore, the science center for SDO includes additional hardware resources which consist of:
\begin{itemize}
\item[--]
Two interactive nodes with a total of 16 cores, 64~Gbytes of memory and 24~Tbytes of disk space. These machines are used for development.
\item[--]
Set of virtual machines, which are essentially used to deploy and test preproduction solutions, and also to host the monitoring system.
\end{itemize}

\subsection{System software}

The software system in charge of managing and storing \linebreak HMI and AIA data was developed by JSOC (see e.g.~Bogart 2007). It consists of two software components built around a relational PostgreSQL database:
\begin{itemize}
\item
The Data Record Management System (DRMS), is in charge of data organization and description, and it maintains database tables containing metadata, such as FITS header keywords and their values.
\item
The Storage Unit Management System (SUMS) is responsible for storing data segments (e.g. the physical data in FITS images) on the dedicated storage resources, and uses database tables to keep track of the online location.
\end{itemize}

The DRMS and SUMS act in a complementary manner to maintain a connection between header information and corresponding data segments.
NetDRMS (figure 3) is a specific version of DRMS/SUMS system that includes a mechanism of sharing metadata and transporting data segments amongst geographically distant sites. The transfer procedure uses a modified Secure Shell (SSH) protocol called High Performance SSH/SCP. NetDRMS is fully operational at the GDC-SDO.

In order to combine large volume of data and complex scientific analysis procedures, we need to orchestrate the scientific modules, the processing tasks, and the data movements. The Pegasus Workflow Management System (WMS) is used to automate pipeline creation and execution. Developed at the University of Southern California Information Sciences Institute, Pegasus (Deelman, et al. 2007) enables constructions and mapping of complex scientific workflows onto local or geographically distributed resources (Grid).

For local execution, the WMS submits jobs to the local Resource Management System (RMS).  The role of the RMS (figure 3) is to ensure optimal use of the cluster.  In our case we have implemented a mechanism based on: Torque/Maui and Condor. The selection between these two mechanisms is made during the workflow creation and depends on the granularity and the degree of parallelism of the scientific modules.

A monitoring system, built around two open source \linebreak tools, Ganglia and Nagios, is used to ensure maximum system infrastructure reliability. Ganglia collects data information and charts several system variables. This information helps us to scale the system needs according to usage. Nagios monitors the entire network infrastructure and provides automatic notification alerts about hardware and services. This allow fast problems resolution.

\section{Conclusion}
SDO has launched and is currently undergoing a commissioning phase where, thus far, all mission operations have been completed successfully. The GDC-SDO is ready to receive HMI and AIA data in the coming months as they become available. An artificial HMI data series, manufactured and distributed by JSOC, was used to validate and verify the automatic transfer software (netDRMS), as well as to ensure that the GDC-SDO has sufficient network capabilities. For the short-term full cadence AIA data, LTO-4 tapes may be posted to Stanford and will remain on standby, ready for when the data comes online. The SDO workflow, and pipeline modules, are under development and are being tested with MDI data. Improvements in the methods of data analysis and their implementation will be required to extract the most scientific insight from the observations.

\acknowledgements
We gratefully acknowledge our collaborators at Stanford University (SDO JSOC), in particular Rick Bogart for his help in setting up NetDRMS at MPS. We would like to thank Ewa Deelman, Gaurang Mehta and Karan Vahi, at USC's Information Sciences Institute (Pegasus project) for their assistance and support with Pegasus and workflow creation. The German Science Center for SDO is supported by a grant from the German Aerospace Center (DLR) and Starting Grant \#210949 from the European Research Council under the European Community's Seventh Framework Programme (FP7/2007-2013).





\begin{thebibliography}{}


  \bibitem{} Birch, A.C., Gizon, L.: 2007, AN~328, 228
  \bibitem{} Bogart, R.S.: 2007, AN~328, 352
  \bibitem{} Cameron, R., Gizon, L., Duvall Jr., T.L.: 2008, SoPh.~251, 291
  \bibitem{} Cameron, R., Gizon, L., Schunker, H.: 2010, arXiv:1003.0528v1
  \bibitem{} Deelman, E., Blythe, J., Gil, Y., et al.: 2003,  JOGC~1, 25
  \bibitem{} Jackiewicz, J.,  Gizon, L., Birch, A.C.: 2008, SoPh.~251, 381
  \bibitem{} Kosovichev, A.G.: 2007,  AN~328, 339
  

\end{thebibliography}
\end{document}